# Edge-pinning effect of graphene nanoflakes sliding atop graphene

Yingchao Liu[1,3], Jinlong Ren[1,3], Decheng Kong[1,3], Guangcun Shan[2,*], Kunpeng Dou[1,3*]

[1]College of Physics and Optoelectronic Engineering, Faculty of Information Science and Engineering, Ocean University of China, Qingdao 266100, China.

[2]Institute of Precision Instrument and Quantum Sensing, School of Instrument Science and Opto-electronics Engineering, Beihang University, Beijing 100191, China.

[3]Engineering Research Center of Advanced Marine Physical Instruments and Equipment of Education Ministry of China & Key Laboratory of Optics and Optoelectronics of Qingdao, Ocean University of China, Qingdao 266100, China.

*Corresponding authors

E-mail address: Guangcun Shan (gcshan@buaa.edu.cn), Kunpeng Dou (doukunpeng@ouc.edu.cn)

**Abstract**

Edge effect is one of the detrimental factors preventing superlubricity in laminar solid lubricants. Separating the friction contribution from the edge atom and inner atom is of paramount importance for rational design of ultralow friction across scales in van der Waals heterostructures. To decouple these contributions and provide the underlying microscopic origin at the atomistic level, we considered two contrast models, namely, graphene nanoflakes with dimerized and pristine edges sliding on graphene monolayer based on extensive *ab initio* calculations. We found the edge contribution to friction is lattice orientation dependence. In particular, edge pinning effect by dimerization is obvious for misaligned contact but suppressed in aligned lattice orientation. The former case providing local commensuration along edges is reminiscent of Aubry's pinned phase and the contribution of per edge carbon atom to the sliding potential energy corrugation is even 1.5 times more than that of an atom in bilayer graphene under commensurate contact. Furthermore, we demonstrated that the dimerized edges as high frictional pinning sites are robust to strain engineering and even enhanced by fluorination. Both structural and chemical modification in the tribological system constructed here offers the atomic details to dissect the undesirable edge pinning effect in layered materials which may give rise to the marked discrepancies in measured friction parameters from the same superlubric sample or different samples with the same size and identical preparation.

# Introduction

Friction converts useful energy into useless heat. To achieve zero friction between two surfaces sliding against each other for saving energy, the slider is required to be infinite long to form incommensurate contact with substrate [1], namely, structural superlubricity [2]. The practical contact of materials is finite. Thus, it is important to consider the influences from edge, the intrinsic "defect" of deposited system on substrate [3]. The significant contribution of edge to friction has been validated in Kr island / Pb(111) [4] and graphene nanoribbon /Au(111) heterostructures [5, 6]. Undercoordinated bonds at edges endow their extra degree of freedoms. As a result, the flexible edges undergo out-of-plane deformation to form local commensurability during sliding which leading to pinning effects and breakdown of superlubricity [3]. Previous studies have reported that the edge pinning effect is load dependent [7,8] and contact size dependent [9]. To diminish the contact edge effects, hydrogen saturation is generally proposed [10] but the edge contribution to friction persisted at a higher load when hydrogen passivation is present [7].

Very recently, there are two benchmark experimental work to decouple the friction contributions from edge atoms and interior atoms in van der Waals heterostructures [11, 12]. One is achieved by graphite mesas with changing the contact size [11]. The high quality measurements about friction parameter show marked discrepancies among different samples with the same size and identical preparation, even for the statistical determination of the same sample from repeated measurements [11]. The possible source giving rise to the discrepancies is deserved to be explored. The other is realized by pushing the sliding on-top domain from the edge or from the center [12]. The surprising result in this work [12,13] is that interior region in the aligned graphene/BN rather than the edge part dominating the friction behavior. This is contrast to the simulation prediction which found the contribution from the near edge region is at least 150% larger than that from the inner atoms per unit area [14].

Since rigid model could not capture the out-of-plane deformation from edge atoms [15], to address the open questions arising in the two aforementioned experimental work [11,12], we explore the edge pinning effect by investigating the atomic scale friction at the interface between a graphene nanoflake and graphene monolayer based on extensive *ab initio* calculations (more than 140 models with number of atoms in the range between 350 and 550 considered, typical samples shown in Fig. 1). Zero normal load was adopted during sliding, which corresponding to the experimental strategy by pushing the edge of slider [12]. It is generally assumed that increasing load would lead to the enhancement of friction [7-9]. We define the variation of adhesion energy $E_{adh}$ ($E_{adh} = E_{flake+substrate} - E_{flake} - E_{substrate}$) to track the corrugation of sliding potential energy along the sliding pathways. The adhesion energy can be decomposed into two contributions from the non-bonded terms, electrostatic and van der Waal's interactions. Both can be qualitatively traced back to the fluctuation of charge transfer and out-of-plane atomic distortions. Among various edge structure environments, the edge dimerization was commonly observed in graphene [16] and so far as we are aware, the contribution from dimer edge to friction has not been reported previously. Based on discussions with atomic details, we found the dimerized edges are extremely detrimental to hamper the superlubricity in misfit contact and robust against strain and chemical modification by fluorine. Thus, the dynamic and inevitable edge dimerization may be the important source to account for the marked discrepancies of measured tribological parameters [11,14].

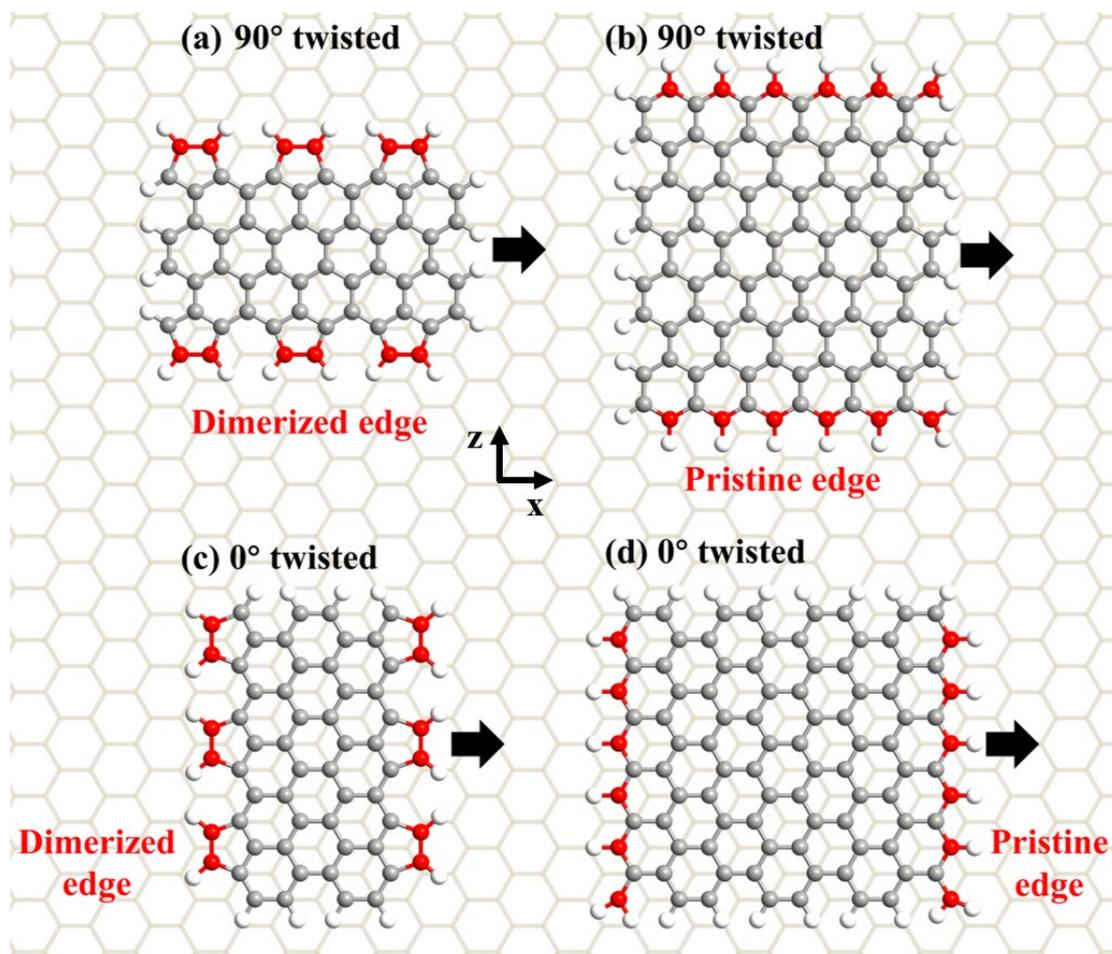

**FIGURE 1 Representative atomic models for sliding simulation.** Misaligned contacts for atop nanoflake with (a) dimerized and (b) pristine edges over graphene substrate. Aligned contacts for atop nanoflake with (c) dimerized and (d) pristine edges over graphene substrate. The edge carbon atoms are highlighted by red color. The black arrows indicate the sliding direction of the nanoflake.

## Results and discussion

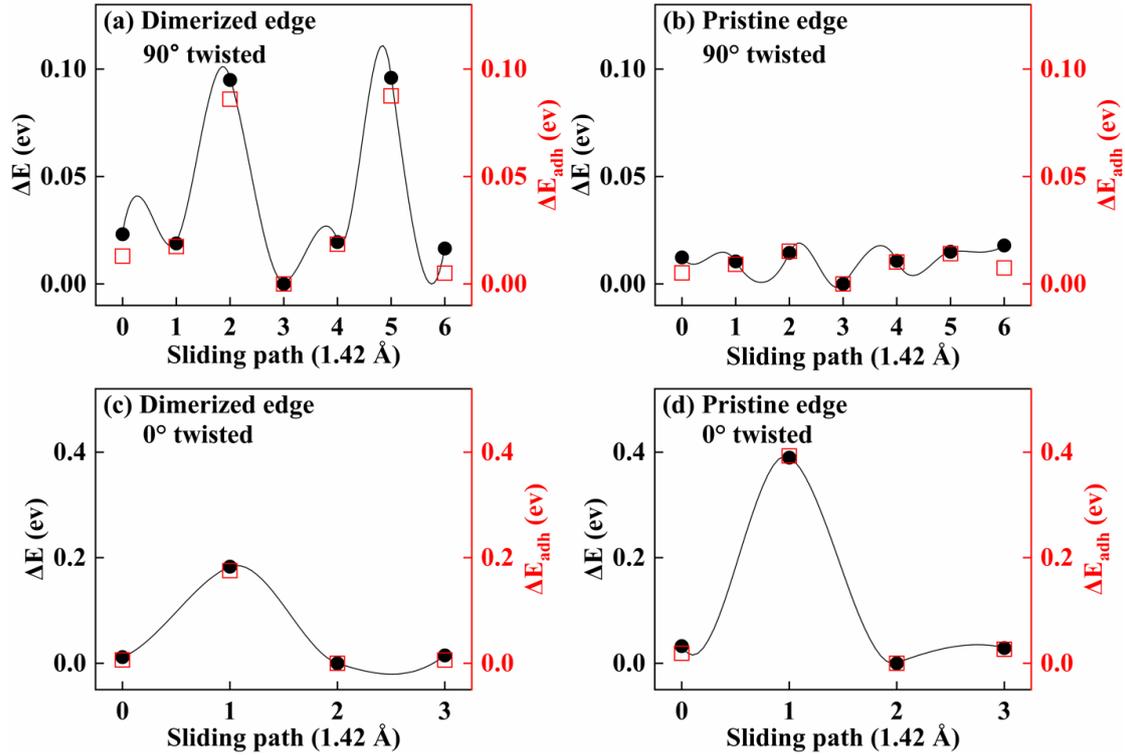

**FIGURE 2 Sliding potential corrugation ΔE versus adhesion energy fluctuation ΔE$_{adh}$.** Correlation between ΔE vand ΔE$_{adh}$ for misaligned contacts of atop nanoflake with (a) dimerized and (b) pristine edges over graphene substrate; aligned contacts of atop nanoflake with (c) dimerized and (d) pristine edges over graphene substrate. The related atomic models are shown in Fig. 1. The adhesion energy E$_{adh}$ at sliding potential energy minimum was set to zero for each case.

## *Dimerized edges acting as high frictional pinning sites in misaligned contact with 90° twisting angle*

Dimerized edges as important source of friction could be understood in the following three aspects. First, as shown in Fig. 2 (a)-(b) (sliding along x axis) and Fig. S1 (a)-(b) (sliding along z axis), sliding barrier is much larger for nanoflake with dimerized edges than that for one with pristine edges over substrate, in spite that the contact area of latter model is 1.72 times larger than that of the former model. The periodical sliding potential profile in Fig. 2 (a) and Fig. S1 (a) reflected the edge pinning effect from local commensurability between the dimerized edges and the substrate, whereas misaligned contact formed between the inner region and substrate. Second, we terminated the nanoflake in Fig. 1(a) by zz(57) reconstructed edge [17] without increasing the contact size too much, as shown in Fig. S2 (a) with four dimers. The ratio of sliding barrier in Fig. S2 (a) and Fig. 1(a) is 0.072 eV / 0.111 eV, close to the ratio of corresponding carbon dimer number 4/6. This is consistent with last aspect, where dimerized edges provide the local commensurability and dominate the friction behavior. Third, we changed the nanoflake in Fig. 1(a) by only top edge dimerized (bottom line with pristine edges) and obtained the sliding barrier, 0.062 eV. The average frictional contribution of a dimer atom could be estimated as 8.2 meV ((0.111 eV–0.062 eV)/6). We also evaluated the average frictional contribution of an atom in bilayer graphene under commensurate contact (zero load) but the value is smaller, 5.4 meV. This indicated that the dimerized edge atoms of graphene nanoflake in misaligned contact contribute even more effectively to friction than those from the bulk counterpart do in aligned interface.

Due to the significant contribution of dimerized edges to the tribological properties, the dynamic and random edge dimerization during sliding may be the important sources to account for the marked discrepancies in measured friction parameters from the same superlubric sample or different samples with the same size and identical preparation [11,14].

We further dissect the atomic origin of pinning effect from dimerized edges. The corrugation of adhesion energy showed good relationship with the fluctuation of sliding potential energy (refer to Fig. 2) and the non-bonded former can be decomposed into two contributions from electrostatic and van der Waal's interactions. Both can be qualitatively traced back to the variation of charge redistribution $\Delta\rho$ and out-of-plane atomic deformations (atomic root mean square displacements, RMSD) as shown in Fig. 3 (a)-(c). With comparison between the sliding potential energy minimum and maximum, the nanoflake with dimerized edge presents obviously fluctuation near edges in both $\Delta\rho$ and RMSD analysis whereas negligible changes in the one with pristine edge. Although the latter model exhibits strong polarization near the edges which is the key factor for the pinning of Kr Monolayers on Pb(111) surface [18], low friction is observed in present work for nanoflake with pristine edge.

## *Friction contributions arising from the edges is suppressed in aligned interfaces with 0° twisting angle*

The nanoflakes with the dimerized and pristine edge both present the periodical sliding potential profiles in Fig. 2 (c) and (d) when slider sitting in the same lattice orientation with respect to the underlying substrate. The sliding barrier of former model is smaller than that of latter model, due to the difference in size of contact area as shown in Fig. 1 (c) and (d). These indicate the inner atoms dominate the friction behavior which provide commensurability between the flake and the substrate. Larger fluctuation of charge density and out-of-plane motion are found in the interior part than those around flake edge, as revealed in Fig. 2 (g)-(i). The suppressed edge contribution to friction may account for the absence of edge pinning effect in aligned heterostructures [12].

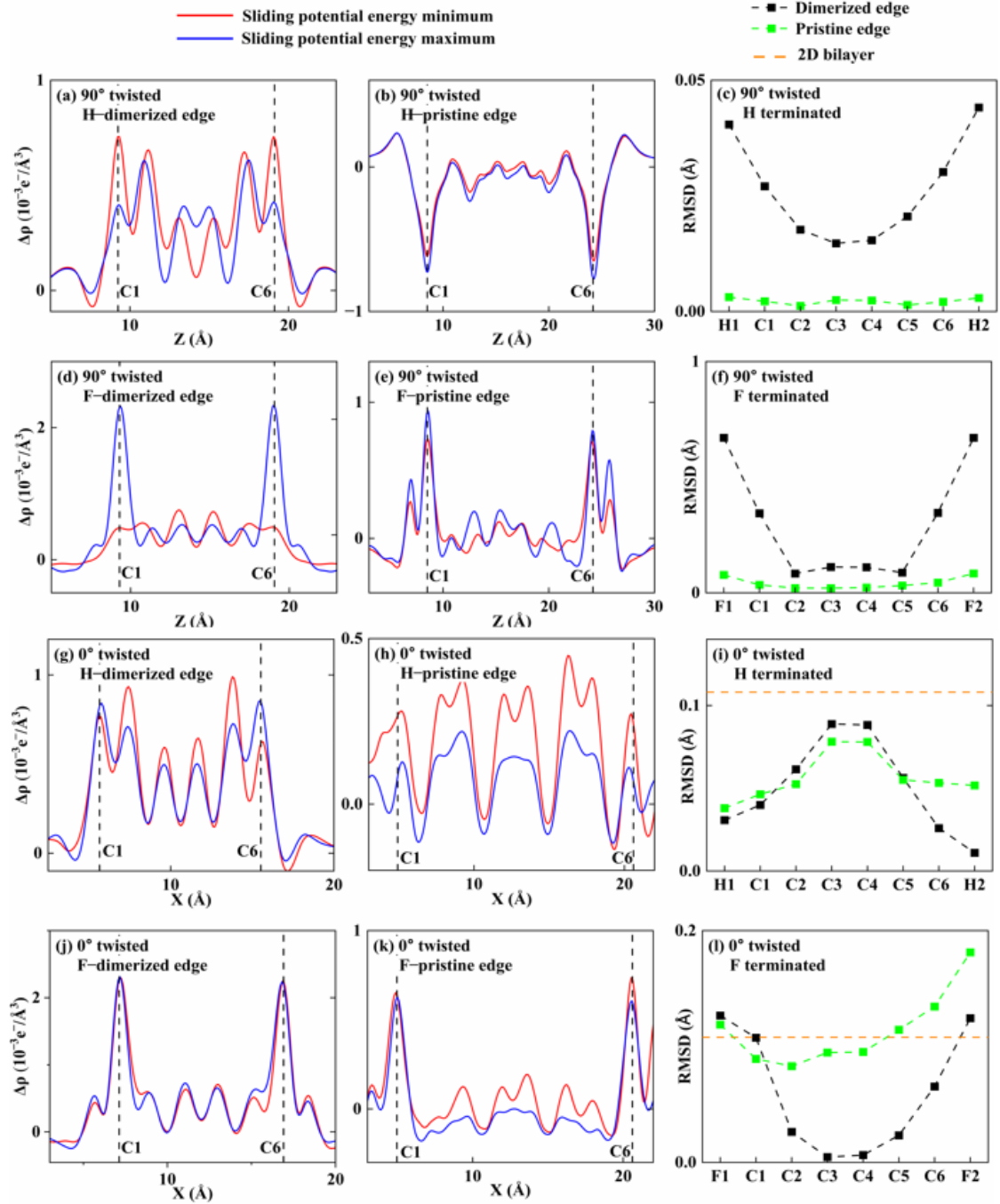

**FIGURE 3 Charge redistribution and out-of-plane atomic deformation for the sliding models.** Electron density differences $\Delta\rho$ along the cutting lines in Fig. S3 are plotted for misaligned contacts: (a) and (b) H terminated edges; (d) and (e) F terminated edges; aligned contacts: (g) and (h) H terminated edges; (j) and (k) F terminated edges. Atomic root mean square displacements (RMSD) are used to address the out-of-plane motion. Misaligned contacts (c) H (f) F terminated edges and aligned contacts (i) H (l) F terminated edges. The horizontal orange dashed line indicates the value of average movements perpendicular to the atomic plane in 2D bilayer counterpart. H1/F1, C1, C2,

C3, C4, C5, C6 and H2/F2 denote the positions for different rows of the nanoflake as marked in Fig. S4.

In the following, we try to diminish the effect of dimerized edges as high frictional pinning sites by two approaches, structural and chemical modification. In particular, we considered strain engineering and fluorinated edges.

*Strain engineering and Aubry's pinned phase*

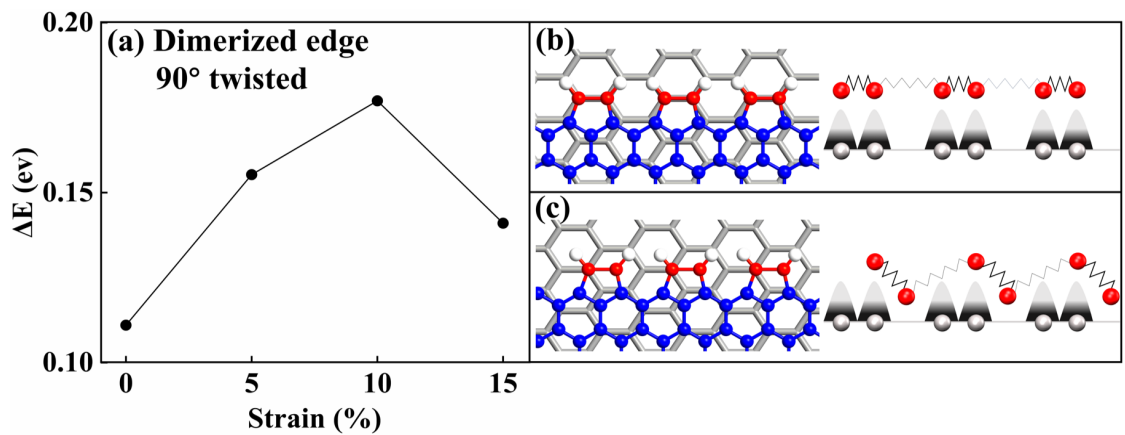

**FIGURE 4 Robustness of edge pinning effect against strain engineering.** (a) Sliding barrier versus uniaxial tensile strain along x axis in misaligned contact for atop nanoflake with dimerized edges. (b) atomic models and (c) schematic view of locally commensurate regions by dimerized edges for sliding barrier maximum at 10% tensile strain. The edge and inner carbon atoms of the flake are colored by red and blue, respectively. The gray lines and balls represent the substrate carbon atoms.

Both experimental [19] and simulation work [20] have successfully demonstrated strain engineering is an effective method to achieve superlubricity in graphene based heterostructures. To break the local pinning capability of dimerized edges, we introduce tensile strain to the substrate along x axis in misaligned contact up to 15%. However, the edge pinning effect persists in the whole range here, reflecting the robustness of dimerized edges acting as high frictional pinning sites. For zero strain, there would be one dimer resting at local energy minima of the substrate one time, hindering relative motion. Whereas for 10% strain (sliding barrier maximum), there would be three dimers simultaneously trapped in local energy minima (Fig. 4 (b) and (c)), which is reminiscent of Aubry's pinned phase [21].

*Fluorinated edges leading to enhanced edge effect*

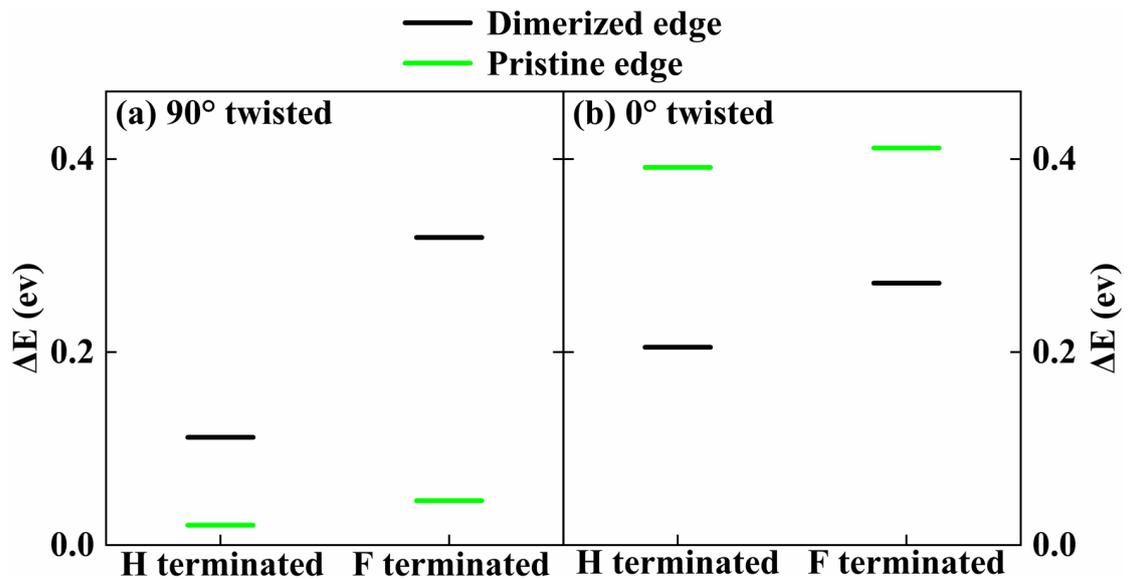

**FIGURE 5 Enhancement of edge pinning effect by fluorinated edges.** Contrasting behavior of sliding barrier for atop nanoflake with H and F saturation at edges in (a) misaligned contact and (b) aligned contact.

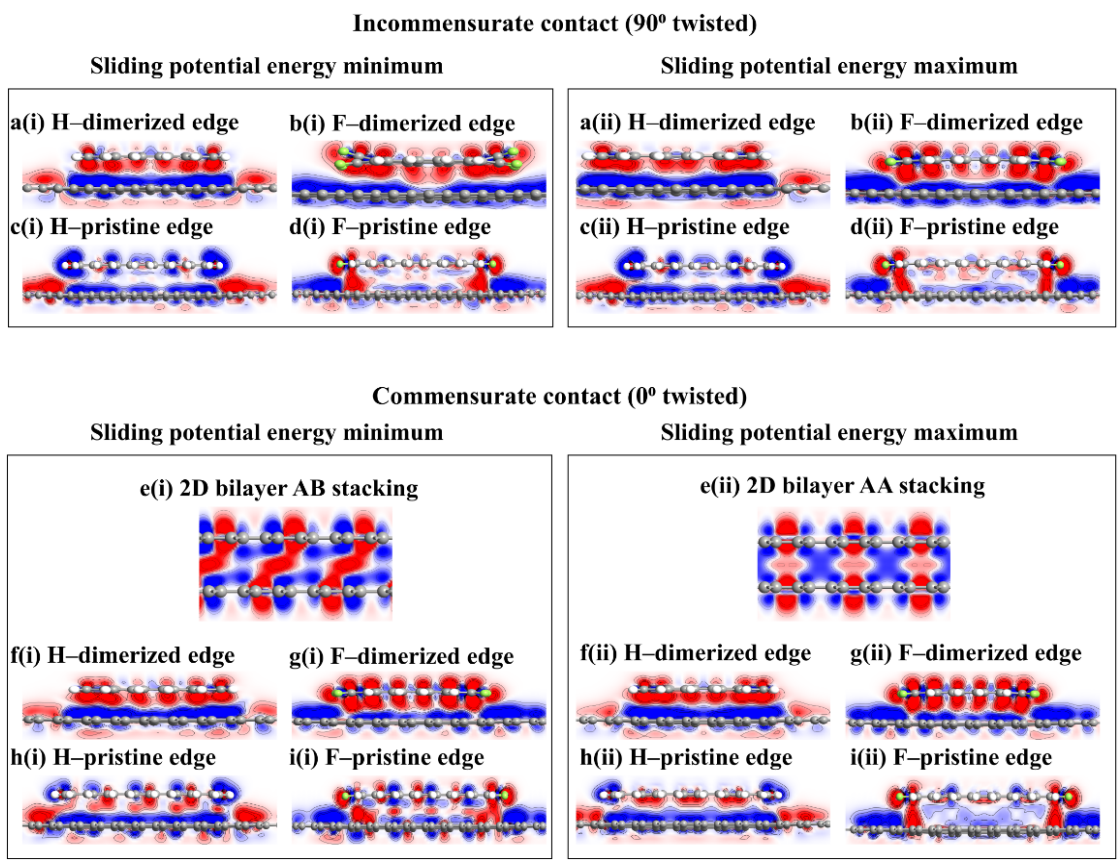

**FIGURE 6 Redistribution of the electron density.** Electron density differences are plotted along the cross sections denoted by the back lines in Fig. S3. Misaligned contacts: (a) H (b) F terminated dimerized edges; (c) H (d) F terminated pristine edges. Aligned contacts: (e) bilayer graphene; (f) H (g) F terminated dimerized edges; (h) H (i) F terminated pristine edges. Red and blue indicate electron accumulation and depletion, respectively.

It has been theoretically anticipated that fluorination of inner carbon atom in commensurate graphene/BN heterostructure plays a key role in reducing the sliding energy corrugation [22]. Fluorinated π-conjugated molecules display strong bending that weakens the electronic coupling between the molecular edges and the substrate [23]. The edge warping of graphene flake could eliminate the edge pinning effect in a very recent experimental work [24]. It is naturally expected that the fluorinated edges would help to achieve structural superlubricity in graphene based heterostructures. Surprisingly, the sliding barrier for all the models here are promoted when changing saturated H to F at edges, regardless of dimerized edge or pristine edge, misaligned or aligned interfaces, shown in Fig. 5. The fluctuation of Δρ are enhanced in misaligned contact but suppressed in aligned contact at edges (Fig. 3 (d), (e), (j) and (k)). The enlarged sliding energy barrier are probably due to the enlarged deformation at edge atoms, irrespective of dimerized or pristine one, as depicted in Fig. 3 (f) and (l).

We further explore the charge transfer analysis in real space (Fig. 6 and Fig. S3), which is commonly adopted in commensurate 2D bilayer heterostructures for tribological properties [25,26]. For the bilayer counterpart here (Fig. 6(e)), superposition of attractive vdW interaction and attractive electrostatic interaction is shown in AB stacking at sliding potential energy minimum ($SPE_{min}$) whereas vdW interaction plus repulsive electrostatic interaction shown in AA stacking at sliding potential energy maximum ($SPE_{max}$). In general, charge transfer in models with dimerized edge provide the simple picture, combination of attractive vdW interaction and attractive electrostatic interaction between slider and substrate, at both $SPE_{min}$ and $SPE_{max}$, irrespective of H or F saturation, misaligned or aligned contact in Fig. 6 (a), (b), (f) and (g). Redistribution of the electron density in models with pristine edge are much more complex: (1) for misaligned interface with H termination in Fig. 6 (c), repulsive electrostatic interaction at both $SPE_{min}$ and $SPE_{max}$; (2) for misaligned interface with F termination in Fig. 6 (d), repulsive/attractive electrostatic interaction at edge/interior part for both $SPE_{min}$ and $SPE_{max}$; (3) for aligned interface with H termination in Fig. 6 (h), the charge transfer is similar to that of the bulk counterpart (Fig. 6(e)) at $SPE_{min}$ but repulsive/attractive electrostatic interaction at edge/interior part for $SPE_{max}$; (4) for aligned interface with F termination in Fig. 6 (i), repulsive/attractive electrostatic interaction at edge/interior part for $SPE_{min}$ whereas repulsive electrostatic interaction at both edge and interior parts for $SPE_{max}$.

## Conclusions

In summary, we have recognized the edge reconstruction for hampering structural superlubricity for the first time. The carbon dimers play as high frictional pinning sites in misaligned contact but suppressed in aligned contact, which may account for the two open questions arising from the recent experimental work [11,12]. We further attempt to diminish the edge effect by structural and chemical modification but both strain engineering to substrate and fluorination at edges verify the robustness of dimers as pinning sites. The models with edge dimerization provide the simple charge transfer picture but present large corrugation of sliding energy, whereas the ones with pristine edge show complex charge transfer accompanied by small fluctuation of sliding energy. The atomic details in present work may help to achieve the superlubricity across the length scales.

## Models and methods

Calculations were performed within Perdew–Burke–Ernzerhof (PBE) density functional [27], as implemented in the Vienna Ab Initio Simulation Package (VASP) [28]. Van der Waals (vdW) corrections were included via the DFT-D3 method [29]. Structural relaxations were allowed until the force acting on each atom was less than 0.02 eV/Å. The space between nanoflakes along the z/x direction was more than 13 Å, which is enough to avoid interaction between the two neighbouring images. More than 20 Å vacuum was used along the non-periodic direction. The Brillouin zone was sampled at Gamma-point. The energy cutoff was chosen as 420 eV. Climbing image nudged elastic band (CI-NEB) method [30] was used to construct the sliding energy pathway.

## Author contributions

## CRediT authorship contribution statement

**YingChao Liu: Jinlong Ren: Decheng Kong:** Data curation, Software. **Kunpeng Dou:** Conceptualization, Data curation, Writing. **Guangcun Shan:** Writing, Supervision, Funding acquisition.

## Data availability

All the data supporting the results of this study are available upon reasonable request to the corresponding author.

## Declaration of Competing Interest

The authors declare that there is no conflict of interest.

## Acknowledgments

This work was supported by Fundamental Research Funds for the Central Universities.

## Appendix A. Supplementary data

Supplementary data to this article can be found online.

# SUPPLEMENTARY INFORMATION

# Edge-pinning effect of graphene nanoflakes sliding atop graphene

Yingchao Liu[1,3], Jinlong Ren[1,3], Decheng Kong[1,3], Guangcun Shan[2,*], Kunpeng Dou[1,3*]

[1]College of Physics and Optoelectronic Engineering, Faculty of Information Science and Engineering, Ocean University of China, Qingdao 266100, China.

[2]Institute of Precision Instrument and Quantum Sensing, School of Instrument Science and Opto-electronics Engineering, Beihang University, Beijing 100191, China.

[3]Engineering Research Center of Advanced Marine Physical Instruments and Equipment of Education Ministry of China & Key Laboratory of Optics and Optoelectronics of Qingdao, Ocean University of China, Qingdao 266100, China.

*Corresponding authors

E-mail address: Guangcun Shan (gcshan@buaa.edu.cn), Kunpeng Dou (doukunpeng@ouc.edu.cn)

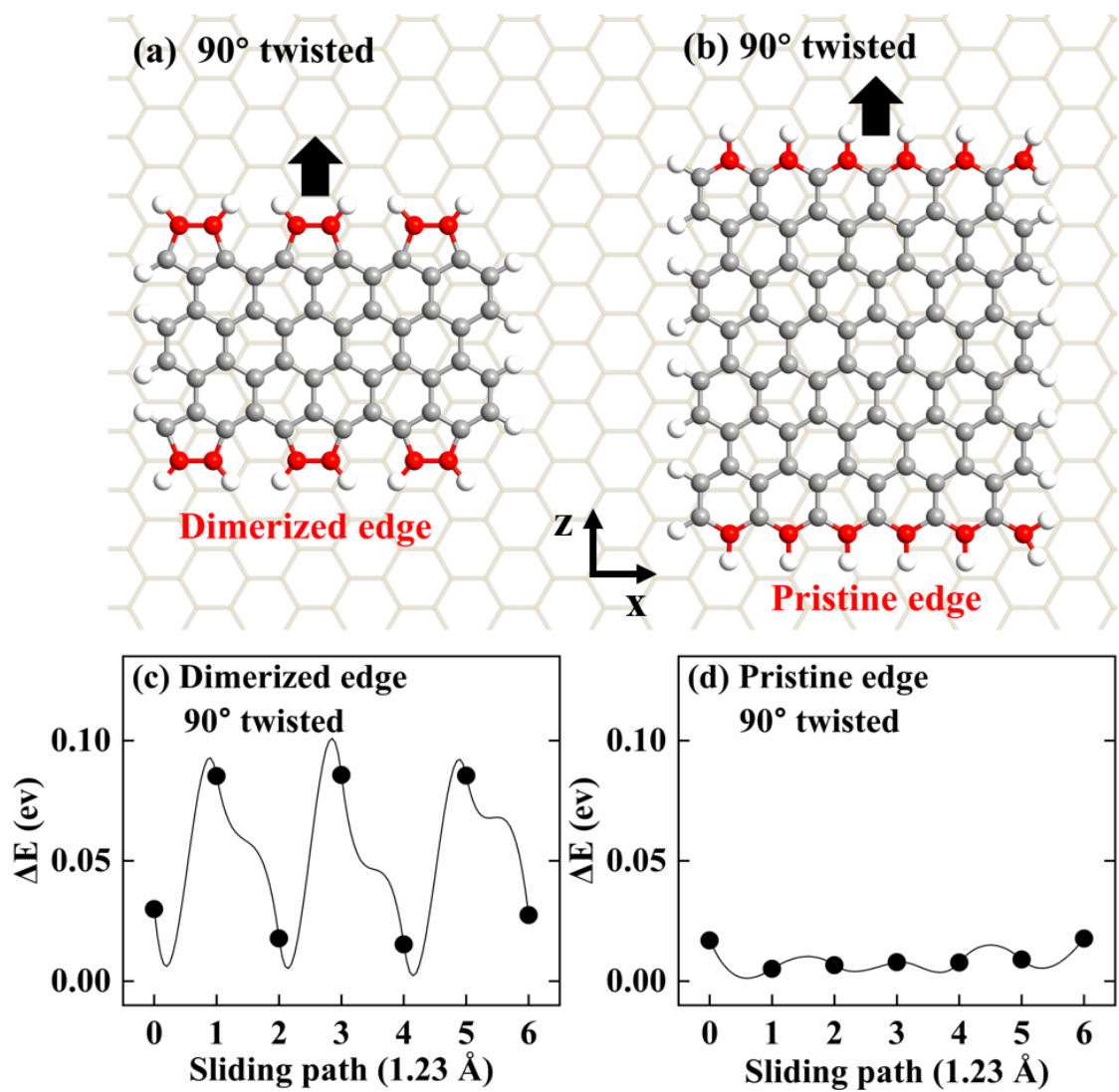

**FIGURE S1** Misaligned contacts for atop nanoflake with dimerized edges and pristine edges over graphene substrate. (a) and (b) atomic models. The edge carbon atoms are highlighted by red color. The black arrows indicate the sliding direction of the nanoflake. (c) and (d) sliding potential corrugation ΔE.

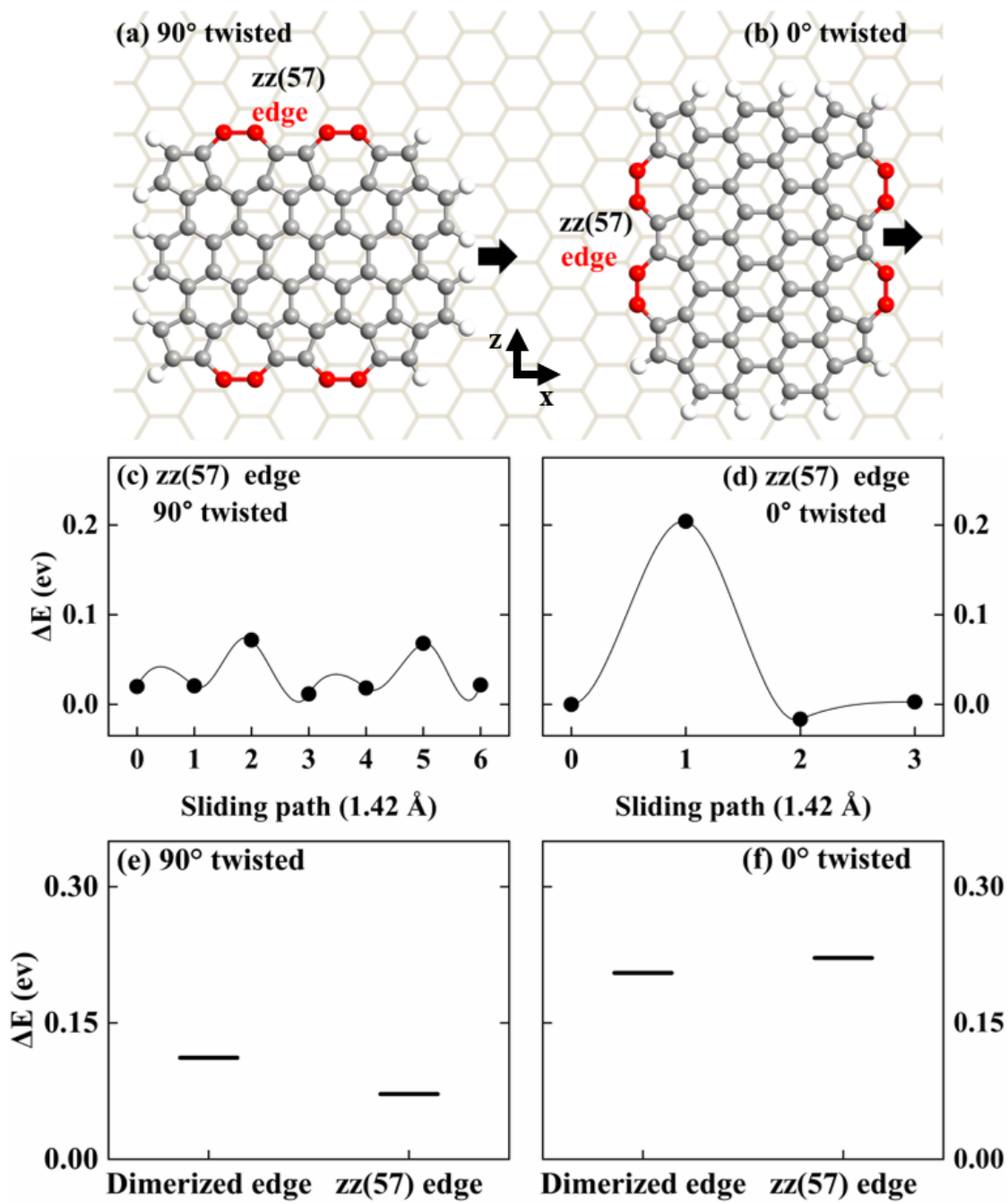

**FIGURE S2** Atop nanoflake with zz(57) reconstructed edges over graphene substrate. (a) and (b) atomic models. The edge carbon atoms are highlighted by red color. The black arrows indicate the sliding direction of the nanoflake. (c) and (d) sliding potential corrugation ΔE for (a) and (b), respectively. (e) and (f) comparison of sliding barrier between the models in Fig. 1 (a)(c) and those in Fig. S2 (a)(b) in misaligned and aligned contact, respectively.

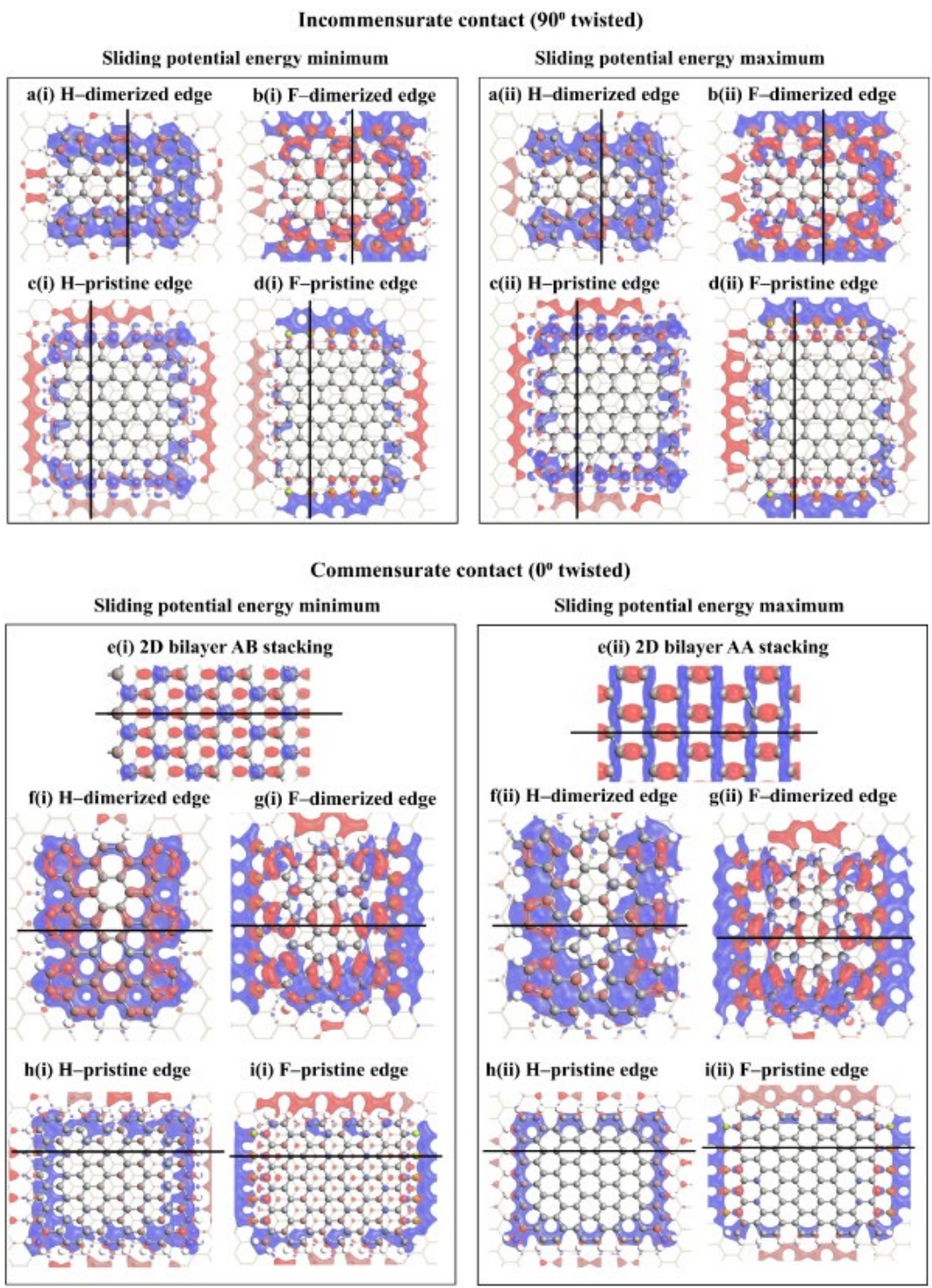

**FIGURE S3** Electron density differences. Misaligned contacts: (a) H (b) F terminated dimerized edges; (c) H (d) F terminated pristine edges. Aligned contacts: (e) bilayer graphene; (f) H (g) F terminated dimerized edges; (h) H (i) F terminated pristine edges. Red and blue indicate electron accumulation and depletion, respectively. The back lines denote the cross sections plotted in Fig. 6.

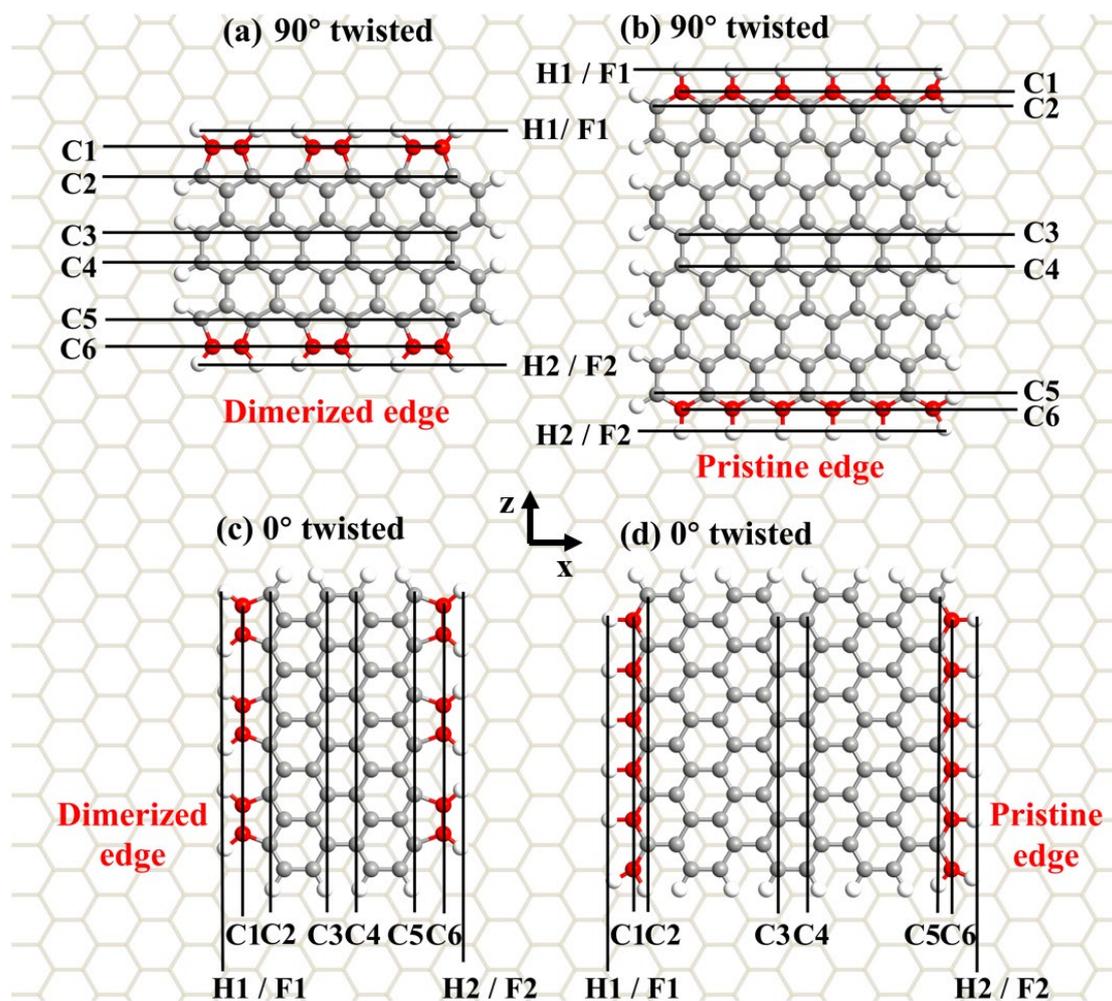

**FIGURE S4** Different rows of nanoflake are defined by H1/F1, C1, C2, C3, C4, C5, C6 and H2/F2, which will be cited in Fig. 3.